\documentclass[sigconf]{acmart}
\usepackage{amsmath,amssymb,amsfonts}
\usepackage{moreverb,algorithm,algorithmic}
\usepackage{graphicx}
\usepackage{caption}
\usepackage{textcomp}
\usepackage{xcolor}
\usepackage{hyperref}
\usepackage{stfloats}
\usepackage{fancybox}
\usepackage{balance}
\usepackage{colortbl}
\usepackage{tabularx}
\usepackage[font=footnotesize]{caption}

\usepackage[normalem]{ulem}
\useunder{\uline}{\ul}{}

\setcopyright{acmcopyright}
\copyrightyear{2019}
\acmYear{2019}
\acmDOI{10.1145/3366486.3366533}
\acmConference[To Appear in WUWNET'19]{WUWNET'19: International Conference on Underwater Networks and Systems}{October 23--25, 2019}{Atlanta, GA, USA}

\begin{document}

\title{UW-MARL: Multi-Agent Reinforcement Learning for Underwater Adaptive Sampling using Autonomous Vehicles}
\author{Mehdi Rahmati, Mohammad Nadeem, Vidyasagar Sadhu, and Dario Pompili}
\affiliation{Department of Electrical and Computer Engineering, Rutgers University--New Brunswick, NJ, USA}
\email{{mehdi.rahmati, mohammad.nadeem, vidyasagar.sadhu,  pompili}@rutgers.edu}

\pagestyle{plain}
\pagenumbering{arabic}
\thispagestyle{empty}

\begin{abstract}
Near-real-time water-quality monitoring in uncertain environments such as rivers, lakes, and water reservoirs of different variables is critical to protect the aquatic life and to prevent further propagation of the potential pollution in the water. In order to measure the physical values in a region of interest, adaptive sampling is helpful as an energy- and time-efficient technique since an exhaustive search of an area is not feasible with a single vehicle. We propose an adaptive sampling algorithm using multiple autonomous vehicles, which are well-trained, as agents, in a Multi-Agent Reinforcement Learning~(MARL) framework to make efficient sequence of decisions on the adaptive sampling procedure.
The proposed solution is evaluated using experimental data, which is fed into a simulation framework. Experiments were conducted in the Raritan River, Somerset and in Carnegie Lake, Princeton, NJ during July 2019. 
\end{abstract}
\maketitle

\section{Introduction}\label{sec:intro}

\textbf{Overview:}
Underwater networks enable various applications such as oceanographic data collection, pollution monitoring~\cite{Sadhu2018ucomms}, and disaster prevention using static nodes~\cite{rahmati2017unisec} and/or mobile vehicles~\cite{rahmati2018probabilistic}. In many applications, we need \textit{mobile vehicles}, equipped with appropriate on-board sensors, rather than static nodes in order to traverse the area and measure the required parameters across the regions of interest, since static nodes pose many limitations~\cite{rahmati2017ssfb} in applications such as adaptive sampling.
%
%
While Remotely Operated Vehicles~(ROVs) are completely human controlled, 
autonomous vehicles, on the opposite end, 
should be able to completely accomplish the required missions by controlling their movement and trajectory without any external inputs. However, using conventional solutions, fully autonomous vehicles are still not fully safe and reliable in navigation and decision-making in practical scenarios. 

\textbf{Motivation and Vision:}
Using one robot to capture the spatial and temporal distributions of the phenomenon with adaptive sampling~\cite{rahmatislam2018} is subject to many constraints, such as a single point of failure and energy inefficiency~\cite{Zhao2018a}. For example, a robot could run out of energy midway during adaptive sampling.
Also, if the robot which performs the sampling fails, the entire data is lost. 
Therefore, in time-critical applications, a multi-vehicle formation is more efficient than a single vehicle mission in terms of energy consumption and processing time~\cite{Sadhu2017icccn}. Using multiple agents leads to less processing time, which in turn corresponds to lower energy consumption. Given the inherent constraints of vehicles, more scheduling and coordination, such as an accurate path planning algorithm, is required to avoid collisions and to maximize the system performance. To reduce energy consumption, we would also like to prevent redundant work done by the vehicles.
%
%
%
To rectify the aforementioned issues, we propose a solution involving multiple agents
in an initially unknown environment that share the workload to perform sampling.
No particular formation is initially enforced by the team. However, the algorithm selects samples such that a desired accuracy in reconstruction is maintained. 


\textbf{Contribution:} 
The goal of this paper is to conduct adaptive sampling using a team of underwater/surface autonomous vehicles using the Multi-Agent Reinforcement Learning~(MARL). 
To achieve this goal, 
the vehicles explore the environment to learn the policy through experience and to make future actions in a MARL framework under uncertainty. The way we handle the problem is different from other non-interactive machine learning problems, since $(i)$~the underwater environment is challenging and impose uncertainties due to the currents; $(ii)$~the communication is performed as long as the vehicles are on the surface; $(iii)$~connection is not possible without acoustic communications and the navigation faces more challenges without the Global Positioning System~(GPS), while the vehicles are underwater; therefore, in our solution, the vehicles surface periodically to reconnect and resume the algorithm; and $(iv)$~a solution is proposed while scalability of the vehicles is considered. We propose map reconstruction and communication protocols that allow coordination between the robots. 
%


\textbf{Related Work:}
Our previous work on distributed adaptive sampling uses multiple gliders to follow a \textit{predefined path} in parallel without communication between the robots~\cite{chen2012distributed}. The gliders traverse the area using a lawnmower-style trajectory. Afterward, they adaptively scan the area by changing the width of the lawnmower trajectory based on the concentration of the valuable data. 
Our other work on underwater adaptive sampling performs a Simultaneous Localization and Mapping~(SLAM) to adjust the trajectory for an underwater robot~\cite{rahmatislam2018}.
%
%
%
A cooperative approach to MARL for autonomous path planning is discussed in~\cite{megherbi2012}. 
Every spot, visited by a vehicle, is marked to reduce the reward for visiting that spot again to prevent redundant work.
In~\cite{zhoushen2011}, estimation of teammates behavior for MARL is proposed for communication constrained scenarios by having each agent store a teammate model for all of its teammates and continuously update this model. Deep Learning is used for collision avoidance and path planning in~\cite{chenliu2017icra} for non-communicating agents. In~\cite{Sartoretti2018}, reinforcement learning is used to coordinate between different portions of a serpentine robot. In~\cite{lalim2015}, a combination of reinforcement learning and flocking control is used for predator avoidance for groups of robots. In~\cite{wang2018}, a reinforcement learning algorithm for AUV trajectory planning was proposed. %
%
In~\cite{Sadhu2016icac}, MARL has been developed with distributed Q-learning for coordinated data collection of an emergency situation among bystanders and drones. The proposed approach uses a centralized Q-table, but the agents update the Q-table in a distributed and synchronous manner. Most of the above works are not directly applicable to the underwater environment as mentioned earlier. Our work however differs in that ours is an underwater environment with communication and navigation constraints.

\textbf{Paper Organization:}
In Sect.~\ref{sec:prop_soln}, we first define the problem 
and then present our proposed solution using MARL.
In Sect.~\ref{sec:perf}, we evaluate our approach via simulations and field experiments.
Finally, in Sect.~\ref{sec:conc}, we draw conclusions and 
discuss 
future work.


\section{Proposed Solution}\label{sec:prop_soln}
In this section, we first present the problem statement 
and then present our proposed approach to solve the problem. 

\begin{figure}[!t]
\centering
\vspace{-3mm}
\centering
\includegraphics[width=0.98\columnwidth]{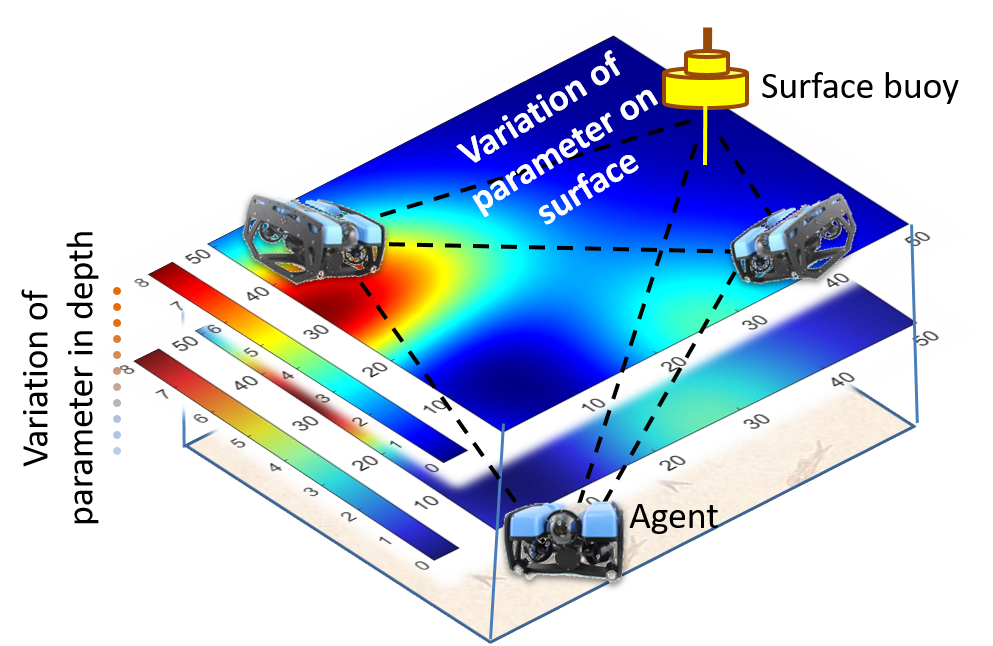}
\vspace{-4mm}
\caption{System model including multiple vehicles, as agents, in a MARL framework. The parameter changes both on the surface and in depth, as shown with the  map.}
\vspace{-3mm}
\label{fig:sysmodel}
\end{figure}

\textbf{Problem Statement:}
%
Assume a set of $V$ vehicles, as agents, in an obstacle-free underwater/surface environment as shown in Fig.~\ref{fig:sysmodel},
which aims to conduct adaptive sampling to measure the variation of environmental parameters, on different locations over the surface and in various depths.      

Reinforcement Learning~(RL), as a model-free data driven approach with a Markov decision process~(MDP) framework, is a novel solution to the problem of multi-vehicle adaptive sampling. Various RL algorithms 
can be utilized based on the application. Q-learning as a value-based algorithm calculates the long-term value
function based on the rewards and actions and is updated step-wise. 
Given all the constraints we mentioned for adaptive sampling, a cooperative multi-vehicle solution under a MARL framework should account to address the aforementioned issues. MARL matches perfectly with Q-learning algorithm and performs by sharing the global Q-function between the vehicles, as agents, or by forming independent Q-learning policies in each vehicle while assuming the other vehicles as part of the system dynamics~\cite{tan1993multi}.

\textbf{Proposed Approach:}
We conduct adaptive sampling in \textit{two phases} through exploration and exploitation of the MARL framework. 
%
%
The region to explore is modeled as a $\mathcal{M} \times \mathcal{N}$ rectangular grid. Each cell on the grid has the same area and dimensions. For each cell, after exploration, we calculate a variance value based on the distribution of the data collected in the same cell, which we associate as the reward for the cell. Exploration phase of MARL is used to obtain this variance data in Phase $1$, through a distributed Q-learning approach~\cite{distqlearn2,distqlearn3}.

The size of the grid is determined by the dimension of the interest region, the battery remaining on the agents' devices and the number of agents available. The \textit{states} are the cells 
of the grid. The \textit{actions} for each state are that of a standard grid-world MDP: \{Left, Right, Up, Down\}. When an agent is given a particular action, it goes to the adjacent cell (up to a specified distance determined by the cell size) in the grid in that particular direction. Whenever an agent takes an action to move to a different state, it captures the data of the current state (grid cell in the MDP) before executing the action. This data determines the \textit{reward} the agent receives for that state. In other words, the reward is dependent on the state alone rather than on the state-action-state sequence. 
A high variance from a cell indicates that the data cannot be predicted in that cell, which means there is more information to gain and hence a higher reward is assigned as $r(s)=Var(\rm{data})$. A more varied dataset implies that searching this cell further is more helpful than searching a cell with more uniform data. For the purposes of reconstructing a map with accuracy, we want to explore the more varied areas. Thus, we want to associate a higher reward for these more varied areas.
Distributed Q-learning method is used where the agents share the same Q-values using a shared database (with synchronous read/write) to expedite the exploration process. 
Each agent still acts independently. As all agents are exploring in parallel and making use of the knowledge/experience gained from all other agents (Q-values), the exploration process is greatly sped up. We use Boltzmann exploration strategy as it is known to give better performance than $\epsilon$-greedy~\cite{Sadhu2016icac}.
In Boltzmann exploration, the probability of selecting an action, $a$, in a given state, $s$, is directly related to its Q-value as 
$\pi_t(s,a) = {e^{Q_t(s,a)/T}}/{\sum_{a'\in A}e^{Q_t(s,a')/T}}$.
Here, the parameter (temperature $T$) decides the amount of exploration or exploitation; if $T$ is large, all actions have almost equal probability resulting in pure exploration whereas when $T$ tends to $0$ the optimal action has the highest probability, resulting in pure exploitation. Hence, it is desired to have a high value for $T$ in the beginning and reduce it gradually to get optimal performance (similar variation is needed for learning rate, $\alpha$).
To enable such variation, the decay in Boltzmann temperature $T$ (same for $\alpha$) is modeled as a radioactive decay, $T=T_{min} +(T_{max} - T_{min})e^{-\frac{\log 2}{tHalf}t}$.

In phase $2$, the vehicles use the data obtained from phase $1$. Each vehicle searches cells based on the expected reward values associated with that location. Vehicles exchange the data through a proposed communication protocol, in which each vehicle broadcasts its decision to go to a certain cell along with the reward associated with that location. Each vehicle has to query the Q-table in order to find the cell which it wants to explore to maximize the rewards. To prevent a cell to be searched twice, a table is stored on the reserved server that keeps track of the searched cells. Before searching a cell, each vehicle queries this table as well, to check the available cells to search. After all cells in the region of interest are explored and exploited, each vehicle shares its map with the others to obtain a global map.

\begin{figure*}[!t]
\vspace{-2mm}
    \centering
    \begin{tabular}{ccc}
\hspace{-2mm} 
    \includegraphics[width=0.23\textwidth,
    height=3.4cm]{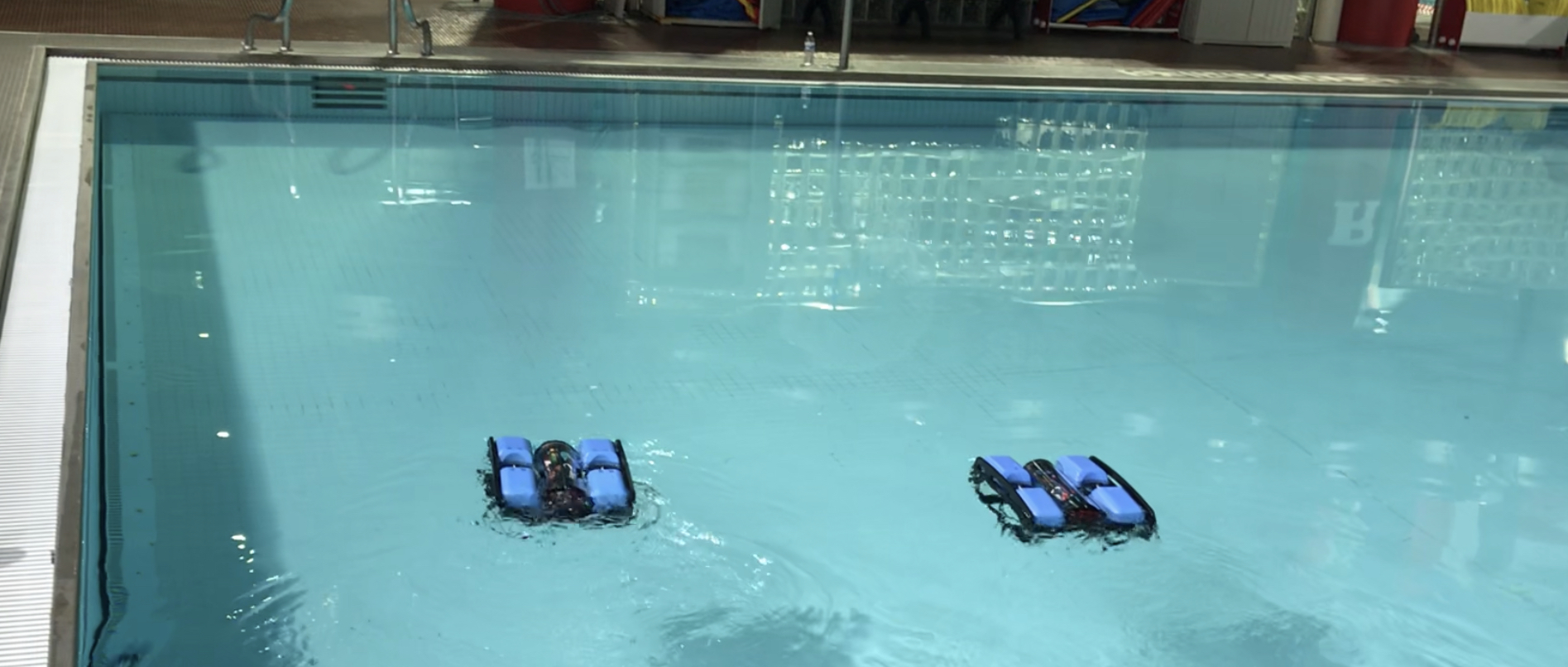}
    \includegraphics[width=0.23\textwidth,
    height=3.4cm]{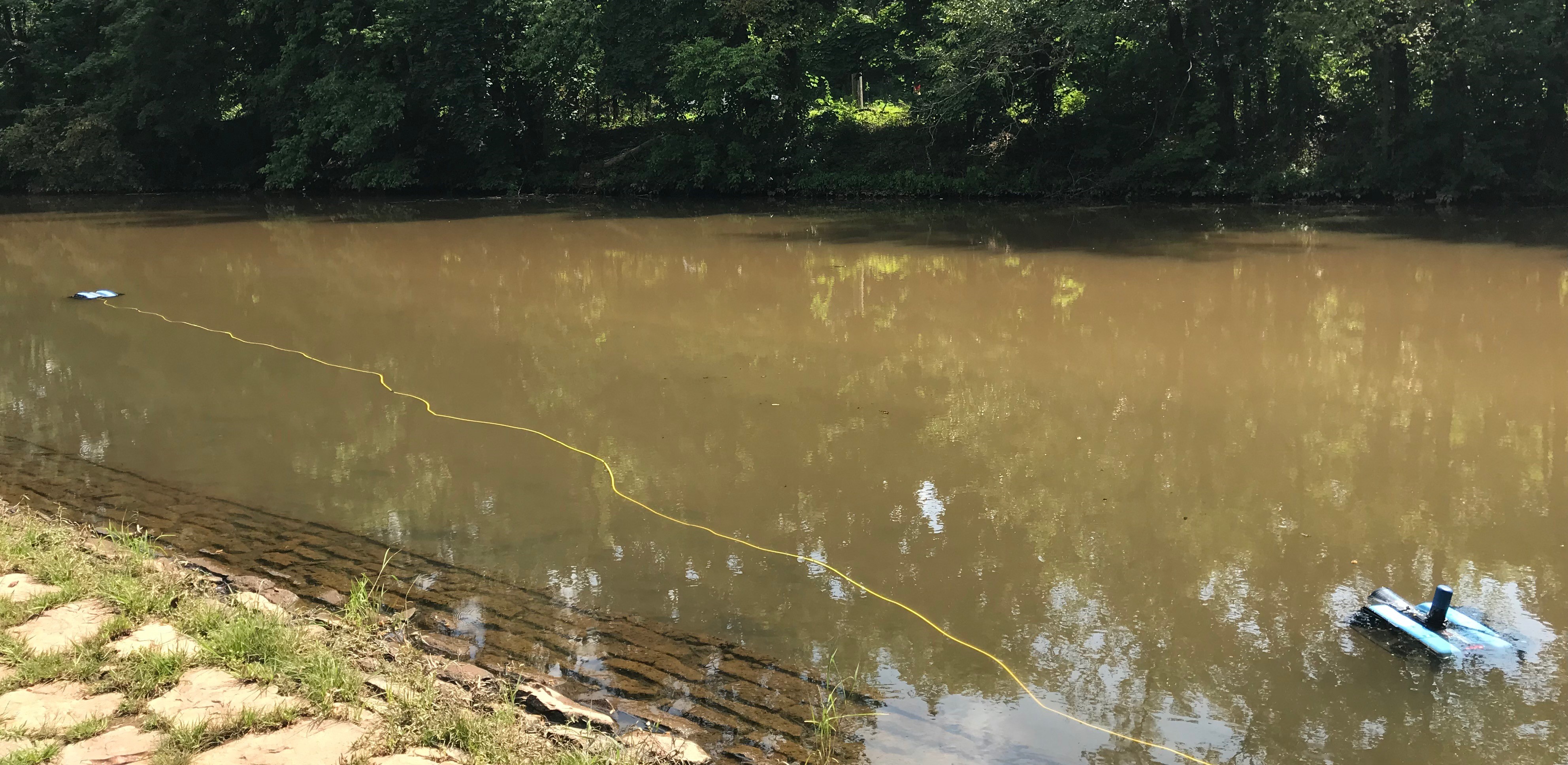}
    \includegraphics[width=0.23\textwidth,
    height=3.4cm]{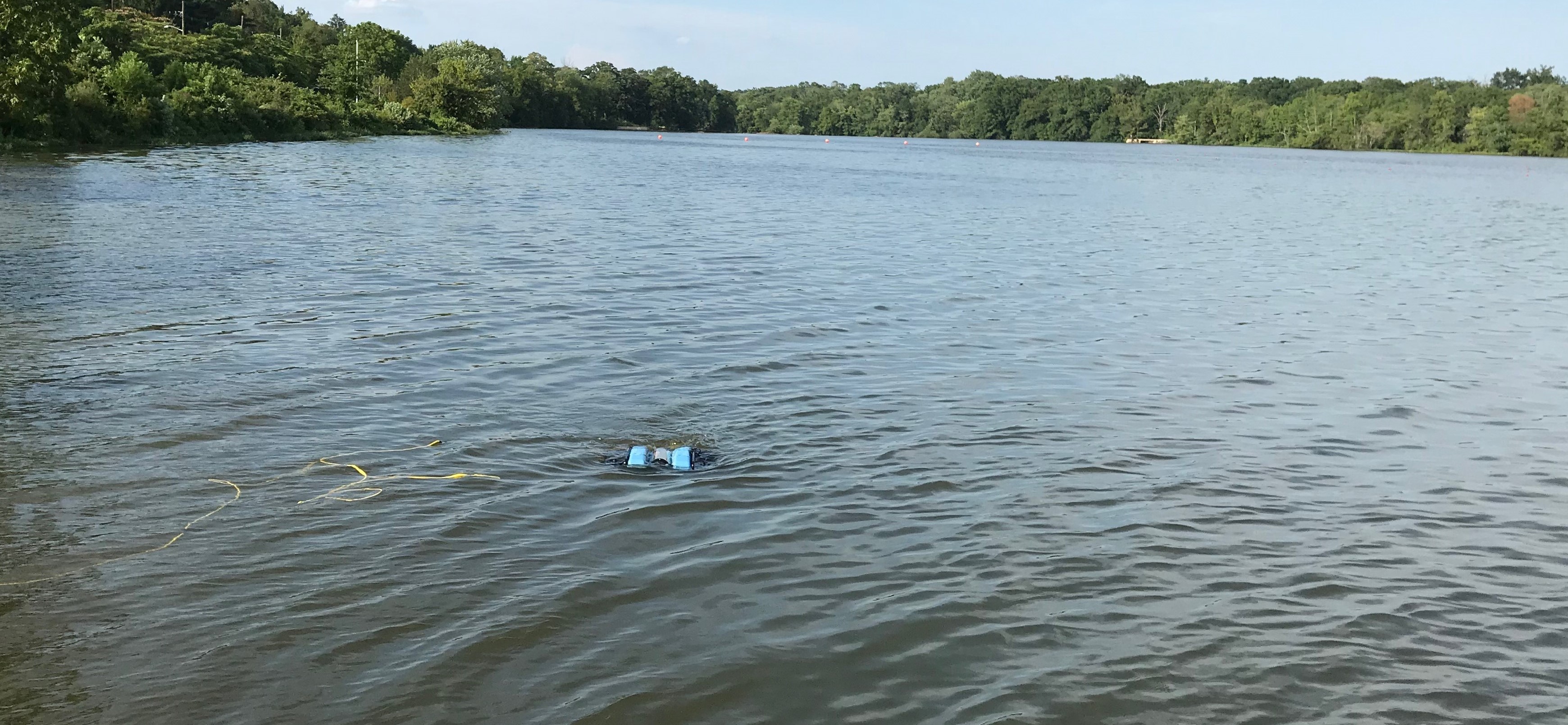}
    \includegraphics[width=0.28\textwidth,
    height=3.4cm]{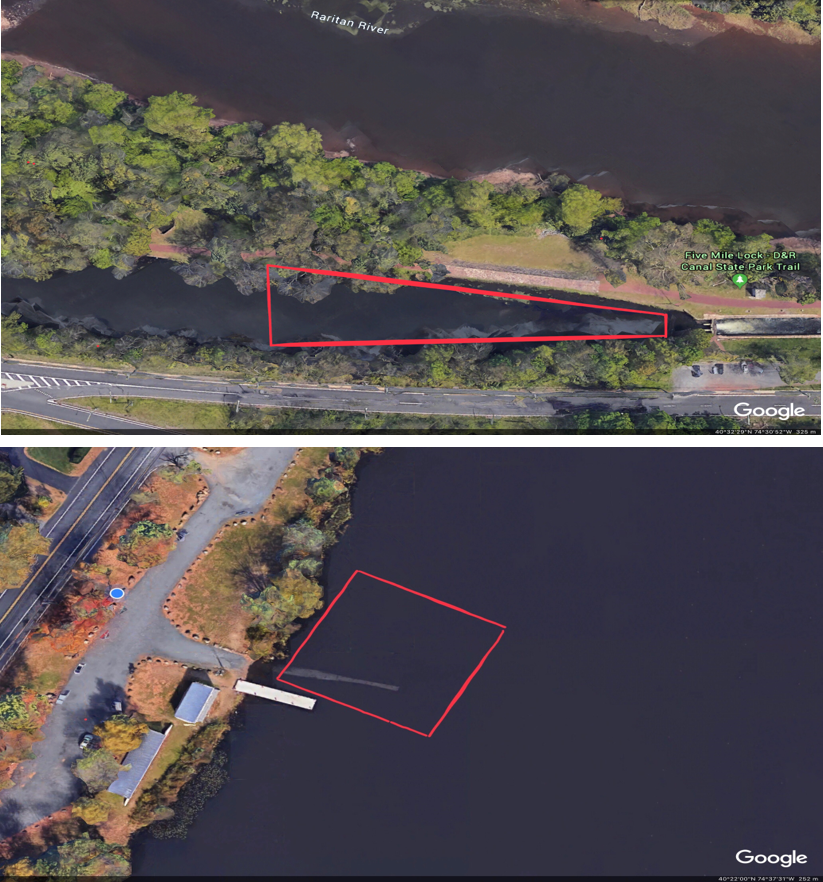}\\
    \vspace{-2mm}
\hspace{-0.5cm} (a)   \hspace{3.5cm} (b) \hspace{4cm}(c) \hspace{3.5cm} (d)
\end{tabular}
    \caption{(a)~The pool at Rutgers Sonny Werblin Recreation Center as a controlled environment to confirm the script written for the autonomous control loop; (b)~Experiments in the Raritan canal in Somerset-New Jersey; and (c)~Experiments in Carnegie Lake in Princeton-New Jersey; (d)~Area of operation: (top)~Raritan canal in Somerset-New Jersey; (bottom)~Carnegie Lake in Princeton-New Jersey.}
    \label{fig:bluerov}
\end{figure*}

\begin{figure}
\vspace{-3mm}
\centering
\begin{tabular}{cc}
\hspace{-3mm}  
\includegraphics[width=4.4cm]{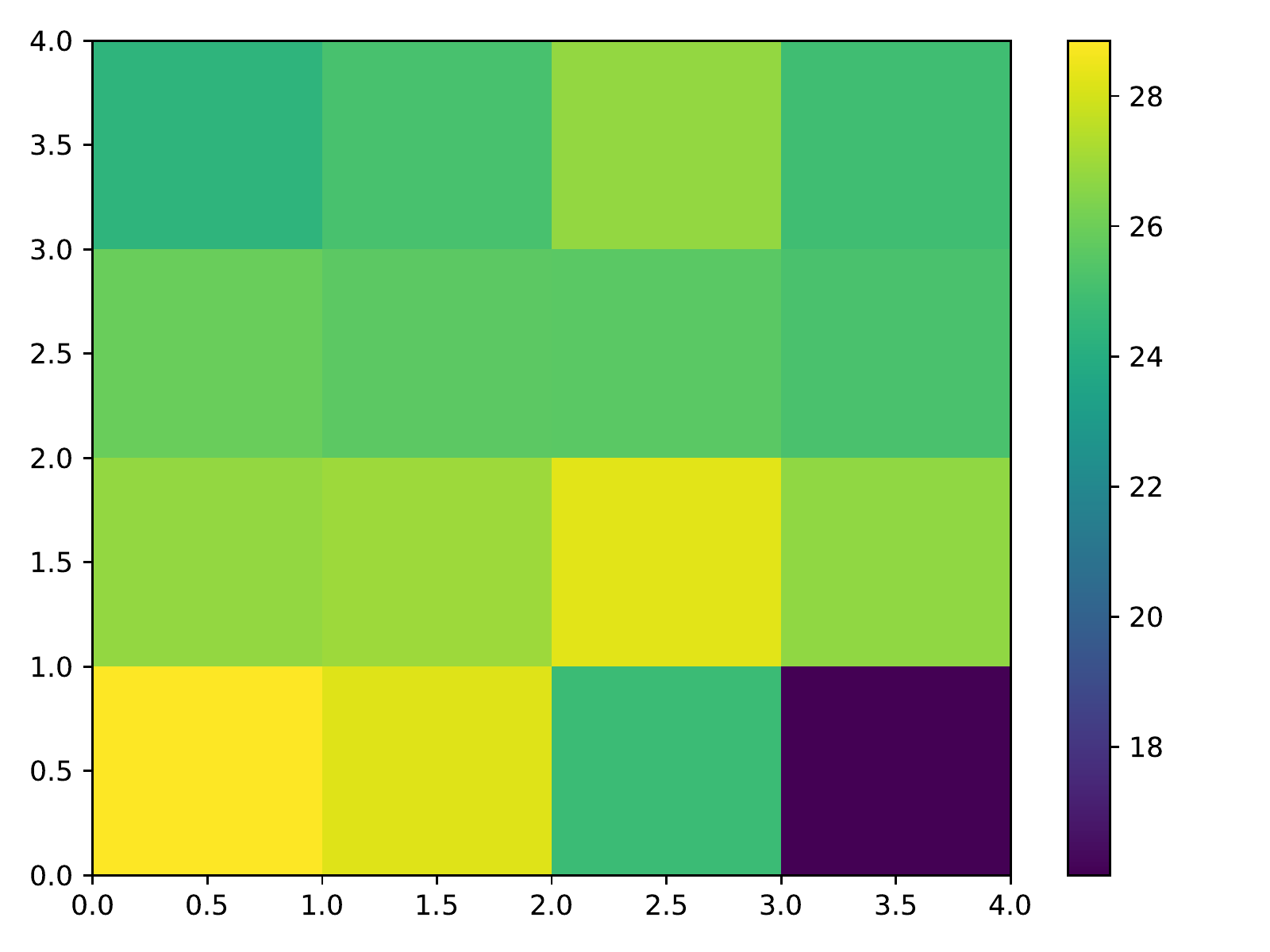}
 \hspace{0.05in}
\includegraphics[width=4.0cm,height=3.1cm]{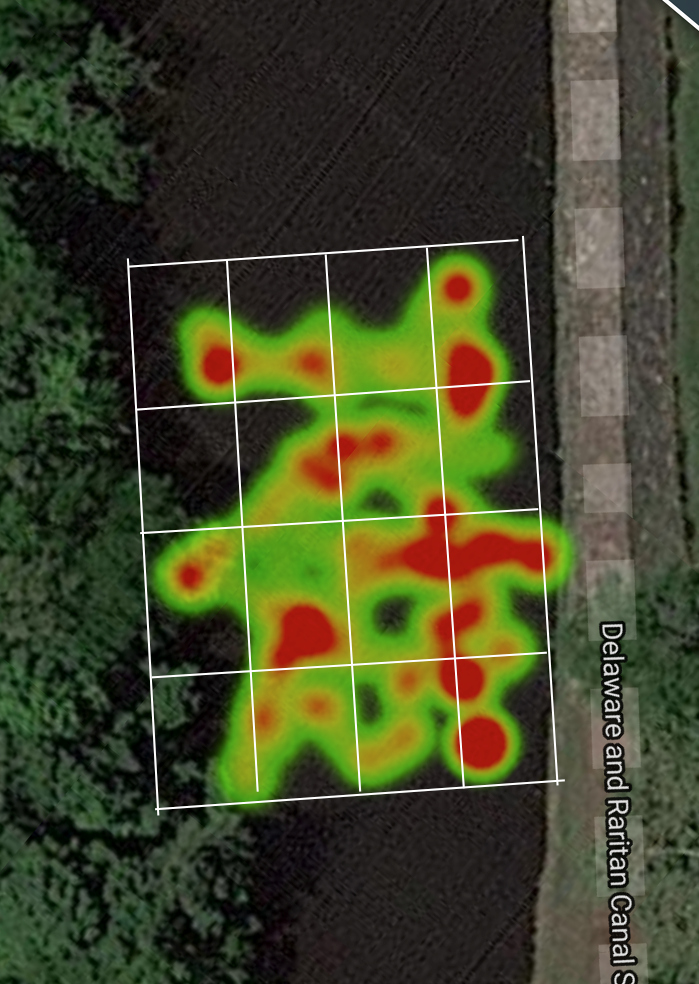}
\\
\vspace{-2mm}
\hspace{-0.5cm} (a)   \hspace{4cm} (b) 
\centering
\end{tabular}
\caption{(a)~Heatmap of the expanded variance data; (b)~Heatmap overlaid on data-collection site (Raritan canal). 
}\label{fig:marlcarnegie}
\vspace{-3mm}
\end{figure}

 






\begin{algorithm}[!t]
\caption{Map Reconstruction.} \label{algo:MapAlgo}
 \small
 \begin{algorithmic}[1] \small
  \REPEAT 
    \STATE{Phase 1 initialization; Grid the area; Communications algorithm( )}
   \UNTIL{receive the data from the vehicles}
    \WHILE{Convergence}
    \STATE{Receive data; Perform MARL algorithm ( )}
    \ENDWHILE
   \STATE{Generate the initial global map; Scan the cells ($C_n, n=1,...,N$)}
   \STATE{Detect the RoIs ($R_i, i=1,...,P , P\leq N$); Find the area of RoI ($D_i$)}
   \STATE{Find the data variation index of each RoI ($\Gamma_i$)}  \%rewards from phase 1
   \STATE{Find the priority index for each ROI ($\Lambda_i=\mu D_i+\beta \Gamma_i$)}
   \STATE{Find the distance between each vehicle ($v_j$) and $R_i$, ($d_{ij}$)}
   \STATE{Sort the $R_i$'s based on $\Lambda_i$}
   \STATE{Sort vehicles based on $d_{ij}+\Psi e_j$}       \,\,\%$e_j$:remaining energy of $v_j$
   \STATE{Initialization for phase $2$: Grid the RoIs}
   \WHILE{Convergence}
    \STATE{Receive data; Perform MARL( ) for each RoI}
   \ENDWHILE
   \STATE{Generate the final global map}
 \end{algorithmic}
\end{algorithm}

\begin{algorithm}[!t]
\caption{Communications Protocol.} \label{algo:commAlgo}
 \small
 \begin{algorithmic}[1] \small
 \STATE{\textit{Asynchronous data communications}}
   \WHILE{Surface MARL mission until convergence }
   \FOR{$i:1:V$}
    \STATE{Broadcast <$ID_{i}$,request>; if no collision: Broadcast <$ID_{i}$,data>}
    \STATE{\textbf{else} Wait for random delay, then resend <$ID_{i}$,request> }
    \ENDFOR
    \ENDWHILE
    \STATE{\textit{Synchronous data communications}}
   \WHILE{Underwater MARL mission until convergence}
    \FOR{$i:1:V$ AND $t:1:\mathcal{T}$}
    \STATE{Surfacing each $t$ second when a cell is finished being explored.}
    \STATE{ID-based Token; Broadcast <$ID_{i}$,data>; All submerge}
    \ENDFOR
    \ENDWHILE
   \end{algorithmic}
\end{algorithm}

Without loss of generality, let a vehicle $v_j$ have the option of searching cells $C_1,C_2...C_N$. The reward associated with each cell is $r_1,r_2...r_N$. Assume that $r_n$ is the maximum reward possible and this reward corresponds to cell $C_n$. However, the vehicle chooses to explore cell $C_m$, where $m\neq n$. As a result, data is not collected from cell $C_n$ which had a more varied dataset and thus, the resultant map is inferior to the map that could be obtained if the vehicle searched cell $C_n$.
For each vehicle, there is a unique pre-defined reward threshold such that any value below this threshold will not be counted as a reward. 
However, for any reward above this threshold, each vehicle will spend time and energy to get it.

\emph{\underline{Map Reconstruction}:}
To reconstruct the global map of environment, the data should be shared among the vehicles following a communications protocol to share the Q-tables. A server is reserved to obtain the data, update the tables, create the map, and send back the updates to the vehicles.
Once the map of the environment is constructed, regions of interest are defined to search further. The definition of these regions varies based on the applications. For example, it could be regions with relatively high values or relatively low values of temperature. MARL framework will rerun for those regions to adaptively sample the field given the constraints occurred in phase $1$ as its inputs. We assume the measured values in the environment changes slowly so that the field between two consecutive rounds of MARL does not experience a considerable change in time and space. Algorithm~\ref{algo:MapAlgo} explains the procedure for two phases of map reconstruction using the MARL framework. After performing the MARL in phase $1$, regions of interest are sorted based on a priority index ($\Lambda$). This index is a scaled combination of the area of the region and its data variation explored in phase $1$ based on weighting coefficients $\mu$ and $\beta$. Then, given the constraints in distance between the vehicles and RoIs, and the remaining energy of each vehicle, the decision is made and an appropriate number of vehicles is assigned to the prioritized RoIs, in which MARL is conducted for a fine reconstruction. The proposed solution is scalable to more number of vehicles.

\emph{\underline{Communications Protocol}:}
The communications protocol is discussed in details in Algorithm~\ref{algo:commAlgo}. As shown here, two different missions are defined: asynchronous surface and synchronous underwater MARL missions. Surface MARL missions use a handshaking, i.e., request to send, when a vehicle is ready to send and to notify others. Since the wireless communications does not work when the vehicles submerge, we define a slotted time window, equal to the maximum required time to surf a grid. Afterwards, the vehicles surface periodically to communicate given an ID-based token.

\section{Performance Evaluation}\label{sec:perf}
We first provide a description of the testbed and experimental setup; then, we provide simulation results using the experimental data.

\textbf{Testbed Description:}
The robot which we exploit is a BlueROV2 (Fig.~\ref{fig:bluerov}(a)) 
and is outfitted with $4$ horizontal and $4$ vertical motors. 
The motor controller
is connected to a Raspberry Pi, 
which is the main computational processor on the BlueROV. We control the motors and receive data through a protocol known as MAVLink.
Sensors interface directly with the Raspberry Pi.
%
We use two BlueROVs with one agent connected to the central server through wired ethernet connection. The other BlueROV is connected through Wi-Fi to the central server and it stays on and scans the surface.

\begin{figure*}
\centering
\begin{tabular}{cc}
\hspace{-3mm} 
\includegraphics[width=4.88cm,height=4.3cm]{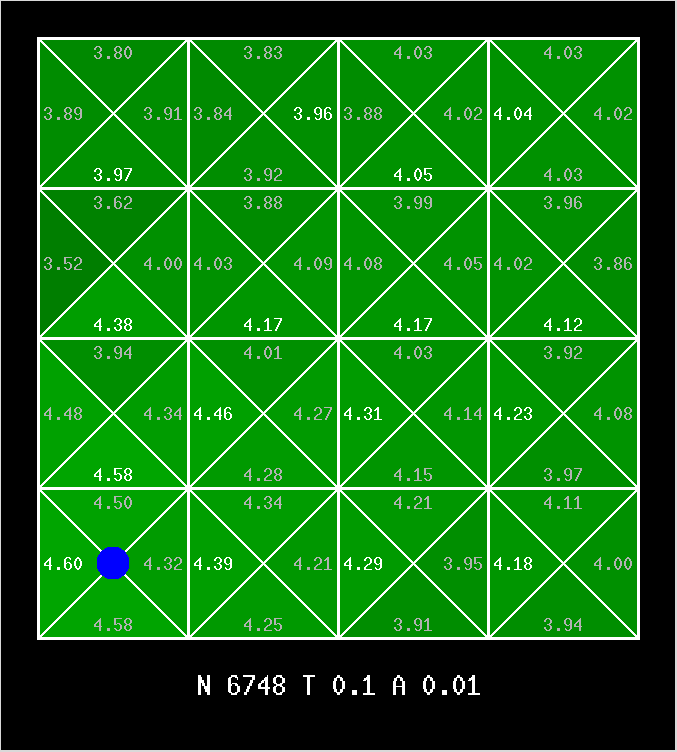}
\hspace{0.05in}
\includegraphics[width=4.88cm,height=4.3cm]{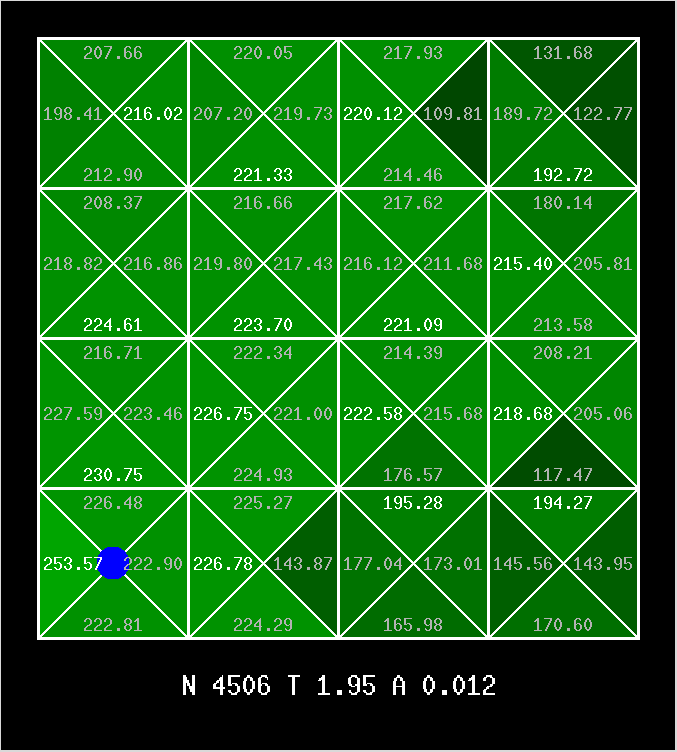}
\hspace{0.05in}
\includegraphics[width=6.6cm,height=4.3cm]{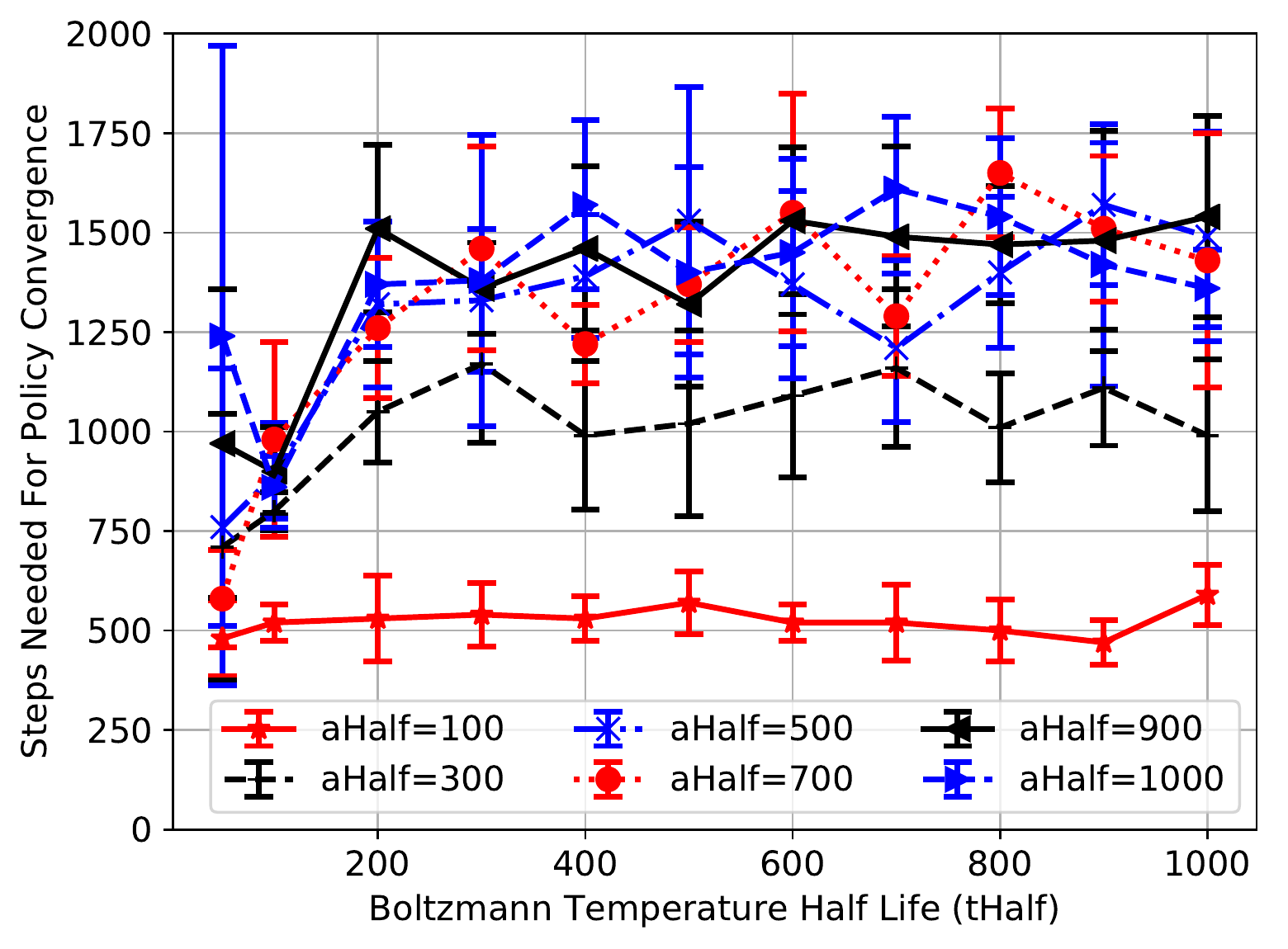}
\\
\vspace{-2mm}
\hspace{-0.5cm} (a)   \hspace{5.5cm} (b) \hspace{6cm}(c)
\end{tabular}
\caption{(a)~
Converged Q-values with actual variance data; (b)~Converged Q-values with expanded variance data---observe convergence in fewer iterations compared to (a); (c)~Number of steps needed for policy convergence for single agent vs. Boltzmann temperature half time~(tHalf) as learning rate half life~(aHalf) is varied.}\label{fig:qvalues}
\vspace{-3mm}
\end{figure*}

\textbf{Experimental Setup and Data Collection:} 
We conducted the data collection procedure in two locations in order to show the varied usability of our approach. The first location was the Raritan River in Somerset, NJ. Here, there were light currents and a mix of shaded areas and exposed areas, which resulted in more variance in the temperature. The second location was Carnegie Lake in Princeton, NJ. Compared to the Raritan, Carnegie Lake had higher currents. The water was completely exposed to the sun. 
Fig.~\ref{fig:bluerov} shows the environments of testing. 
First, we confirm the functionality of the robot in the water as an autonomous vehicle at Rutgers Sonny Werblin Recreation Center as a controlled environment, as shown in Fig.~\ref{fig:bluerov}(a). Two different test locations, as mentioned above, are shown in Figs.~\ref{fig:bluerov}(b)-(c). We split the region into a $4 \times 4$ grid and found the rewards for each grid using our data with dimensions of the region being 8 m $\times$ 8 m. Then, we ran a simulation using the obtained data. The geographical location of the test locations are specified with a red line and shown in Figs.~\ref{fig:bluerov}(d).

\textbf{Experimental Results:}
Fig.~\ref{fig:marlcarnegie}(a) shows the heatmap of the expanded variance data (rewards) collected from Raritan River/Canal. The original variance data had not much difference among the values of different cells. Hence, we expanded the variation by using the following transformation, $100^{\sqrt{x}}$. These values are shown in Fig.~\ref{fig:marlcarnegie}(a). The same heatmap overlaid on the actual experiment location site is shown in Fig.~\ref{fig:marlcarnegie}(b). The data collected varied from 28.3 degrees Celsius to 29.4 degrees Celsius. Experimentally, each iteration takes about 6 seconds to search a grid and to navigate to another grid. Based on Fig.~\ref{fig:qvalues}(c), we require about one hour to search an area effectively for low aHalf and tHalf values for a single agent. Using multiple agents will reduce the time necessary.

\textit{Exploration and Convergence:}
We have applied MARL with distributed Q-learning to the variance data explained above. We used a desktop computer with the MARL algorithm coded in Python to produce these results. We have set $T_{min}=0.01$ and $T_{max}=1000$; $\alpha_{min}=0.001$ and $\alpha_{max}=1$ Fig.~\ref{fig:qvalues} shows snapshots of the exploration for one agent. A lighter green indicates a high Q-value, while a darker (close to black) indicates a lower Q-value. The parameters at the bottom are explained as follows. $N$ indicates the current iteration number; $T$ current Boltzmann temperature which is varied depending on $tHalf$; $A$ denotes current learning rate, which is varied depending on $aHalf$. Fig.~\ref{fig:qvalues}(a) shows the converged Q-values (value convergence) on the original variance data, while Fig.~\ref{fig:qvalues}(b) shows the same on the expanded variance data. We can notice that convergence happens faster on expanded variance data.
Hence for the below results, we have used expanded variance data.

\begin{figure*}
\centering
\vspace{-3mm}
\begin{tabular}{cc}
\hspace{-4mm}  
\includegraphics[width=5.9cm]{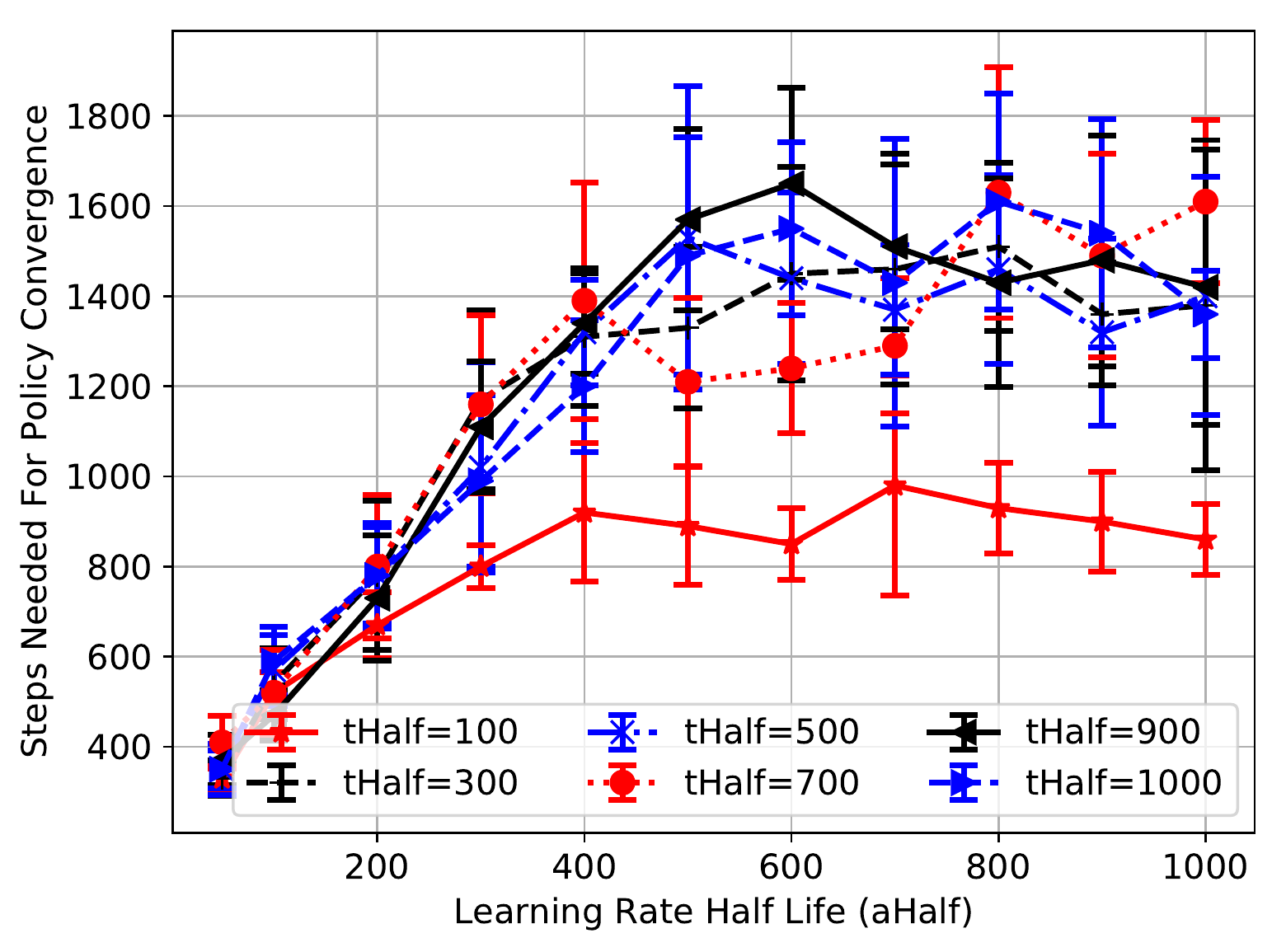}
\includegraphics[width=5.9cm]{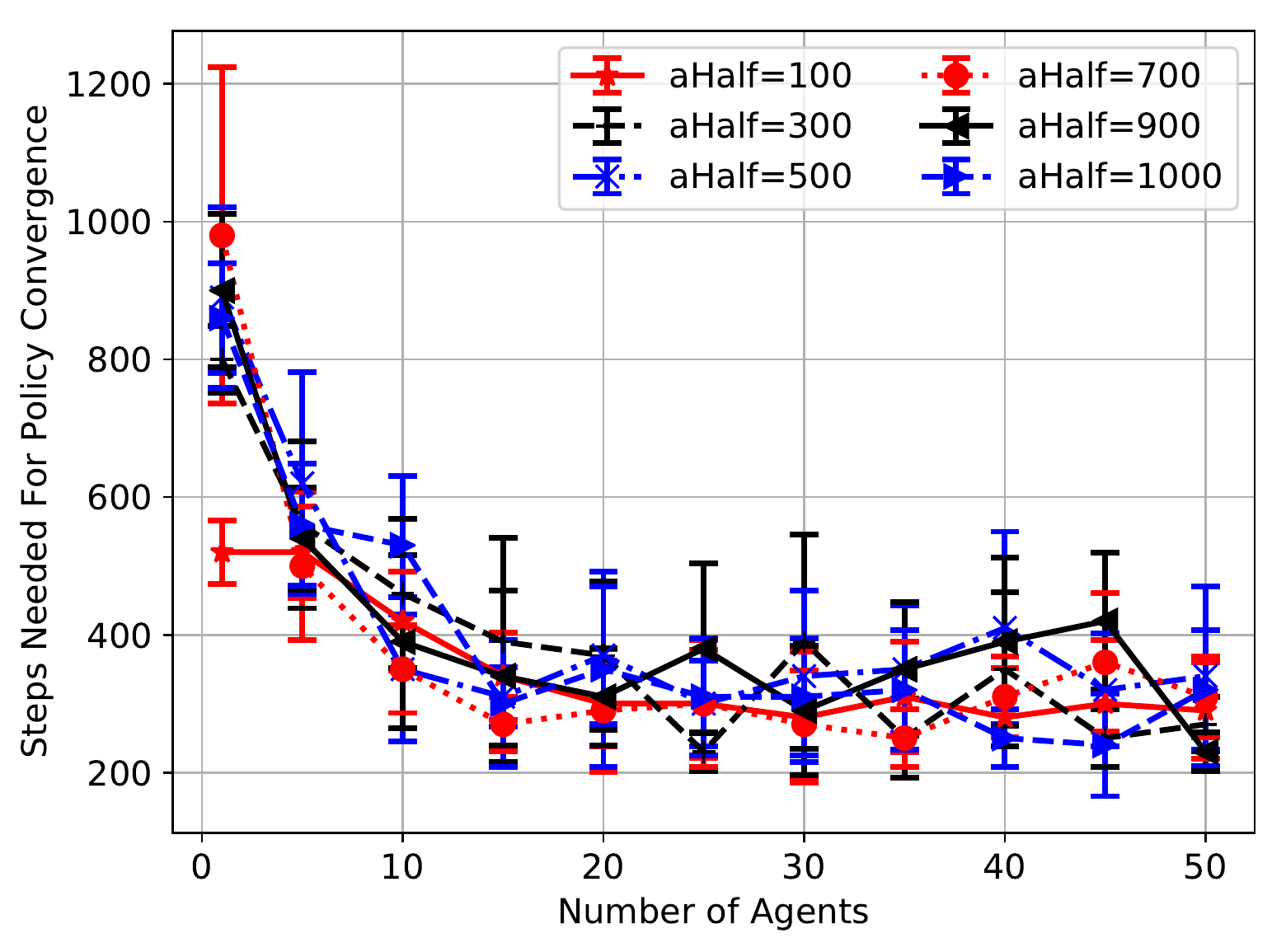}
\includegraphics[width=5.9cm]{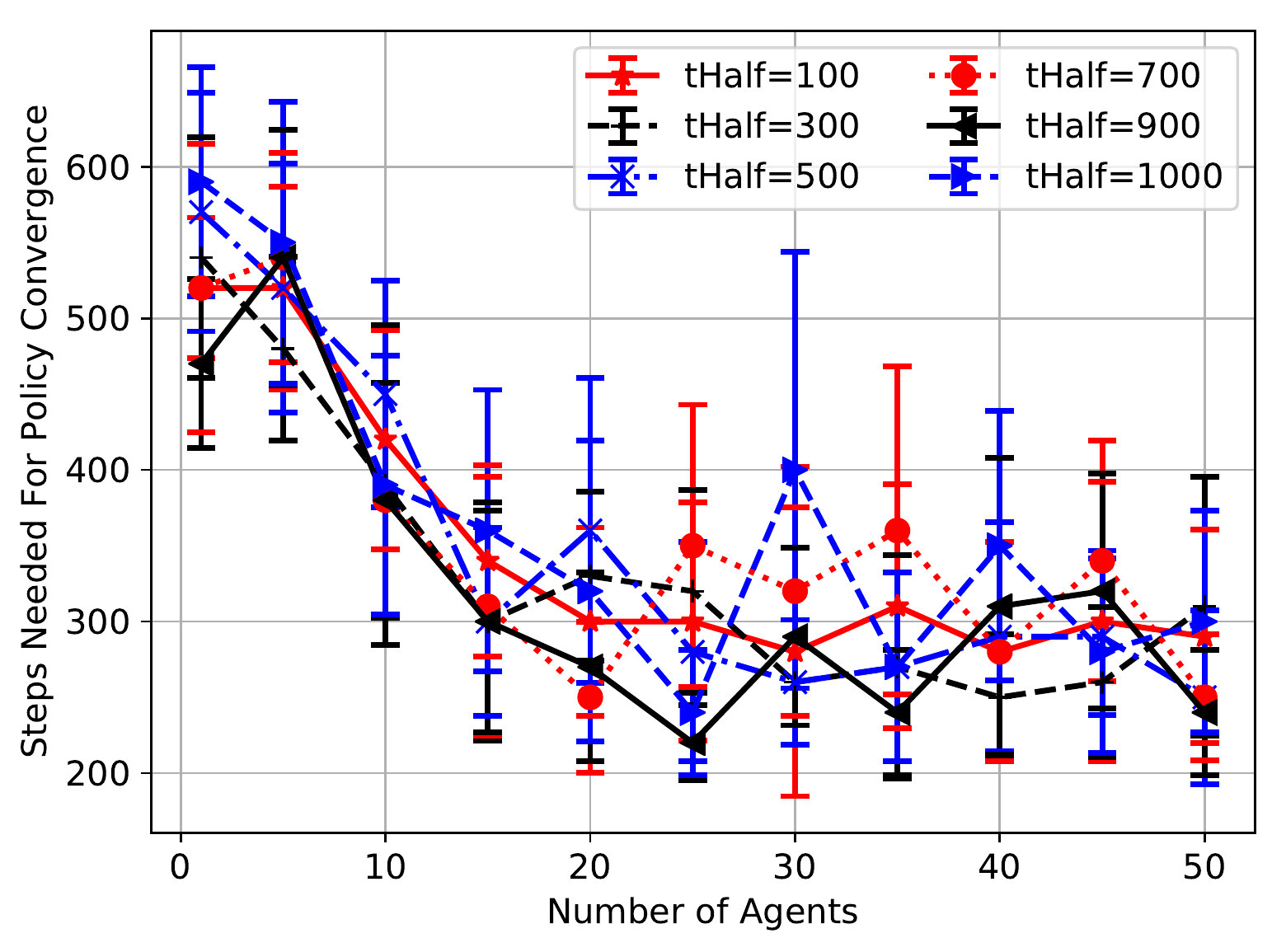}\\
\vspace{-2mm}
\hspace{-0.5cm} (a)   \hspace{4.5cm} (b) \hspace{5cm}(c) 
\end{tabular}
\caption{(a) No. of steps needed for policy convergence for single agent vs. learning rate half time (aHalf) as Boltzmann temperature half life (tHalf) is varied; (b)~No. of steps needed for convergence as number of agents and aHalf are varied; (c)~No. of steps needed for convergence as number of agents and tHalf are varied.}\label{fig:numsteps}
\vspace{-3mm}
\end{figure*}

\textit{Variation with number of agents, aHalf, and tHalf:}
We have studied the policy convergence behavior by varying the number of agents, aHalf, tHalf. Fig.~\ref{fig:qvalues}(c) shows the number of steps required for policy convergence for a single agent when tHalf is varied for different values of aHalf. We can notice, a lower tHalf is preferred, while there is no much variation with higher tHalf values. Also we can observe a lower aHalf is preferred. Fig.~\ref{fig:numsteps}(a) shows the same result when the variation with aHalf and tHalf is interchanged. We can again observe a similar behavior that a lower aHalf and a lower tHalf is preferred. However, we notice that as the number of agents increases, this behavior is no longer observed. Fig.~\ref{fig:numsteps}(b) and (c) show the number of steps needed for policy convergence as the number of agents is varied for different values of aHalf and tHalf. These figures show that the values of aHalf and tHalf effect less with increase in number of agents. This is due to the fact that the exploration is greatly sped up with more number of agents; hence number of agents plays a major role in steps needed for convergence compared to aHalf/tHalf. On the other hand, we observe an interesting trend with the number of agents. We can observe from both figures that the number of steps needed for convergence greatly reduces as the number of agents increases. While this behavior is predominantly seen for number of agents increasing from $1$ to $20$, it diminishes after that. Hence for the given grid size and data, there is no much benefit in having number of agents more than $20$.

\begin{figure*}
\centering
\begin{tabular}{ccc}
\hspace{-3mm}  
\includegraphics[width=4.8cm]{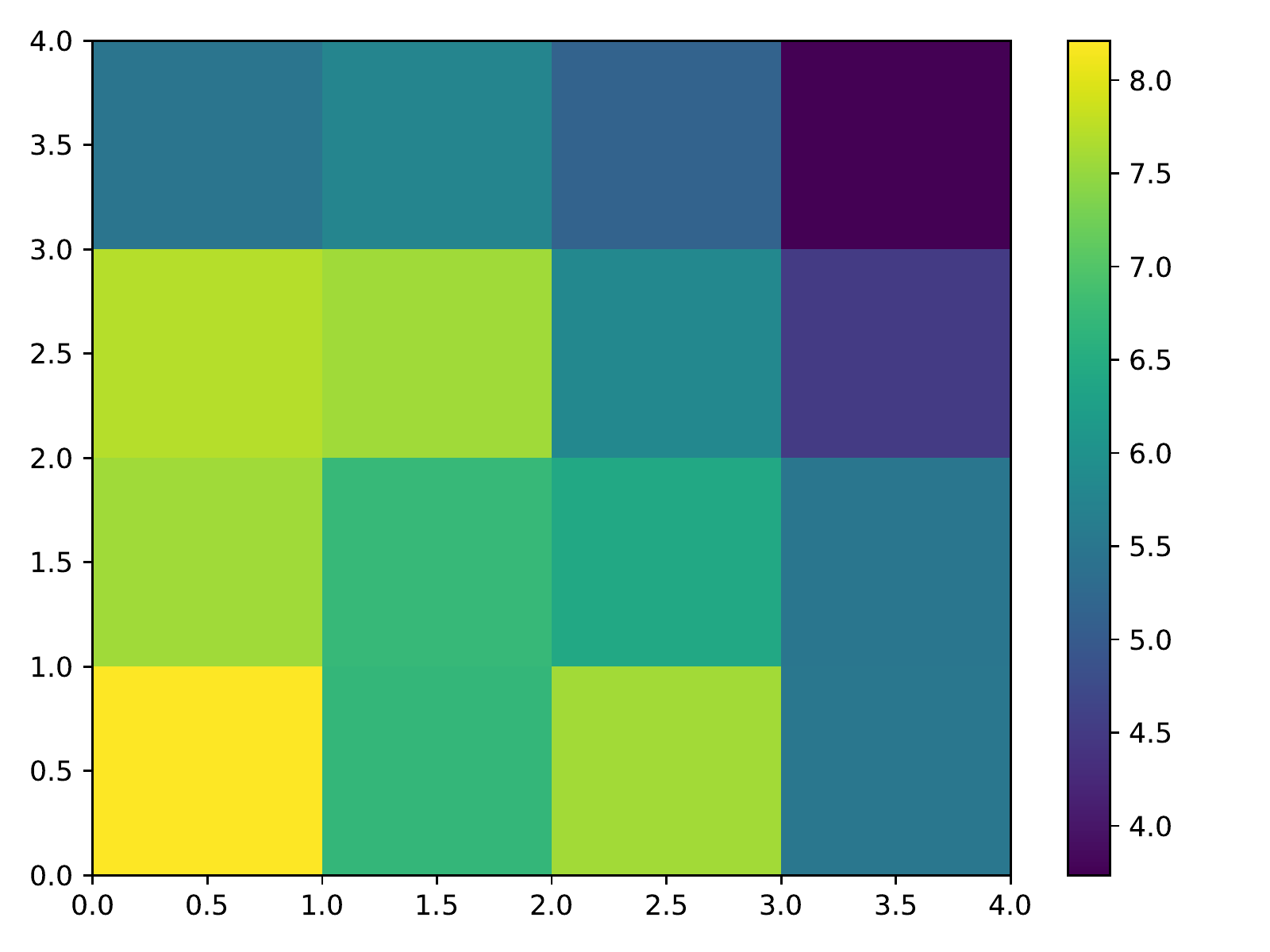}
\hspace{-0.15in}
\includegraphics[width=4.8cm]{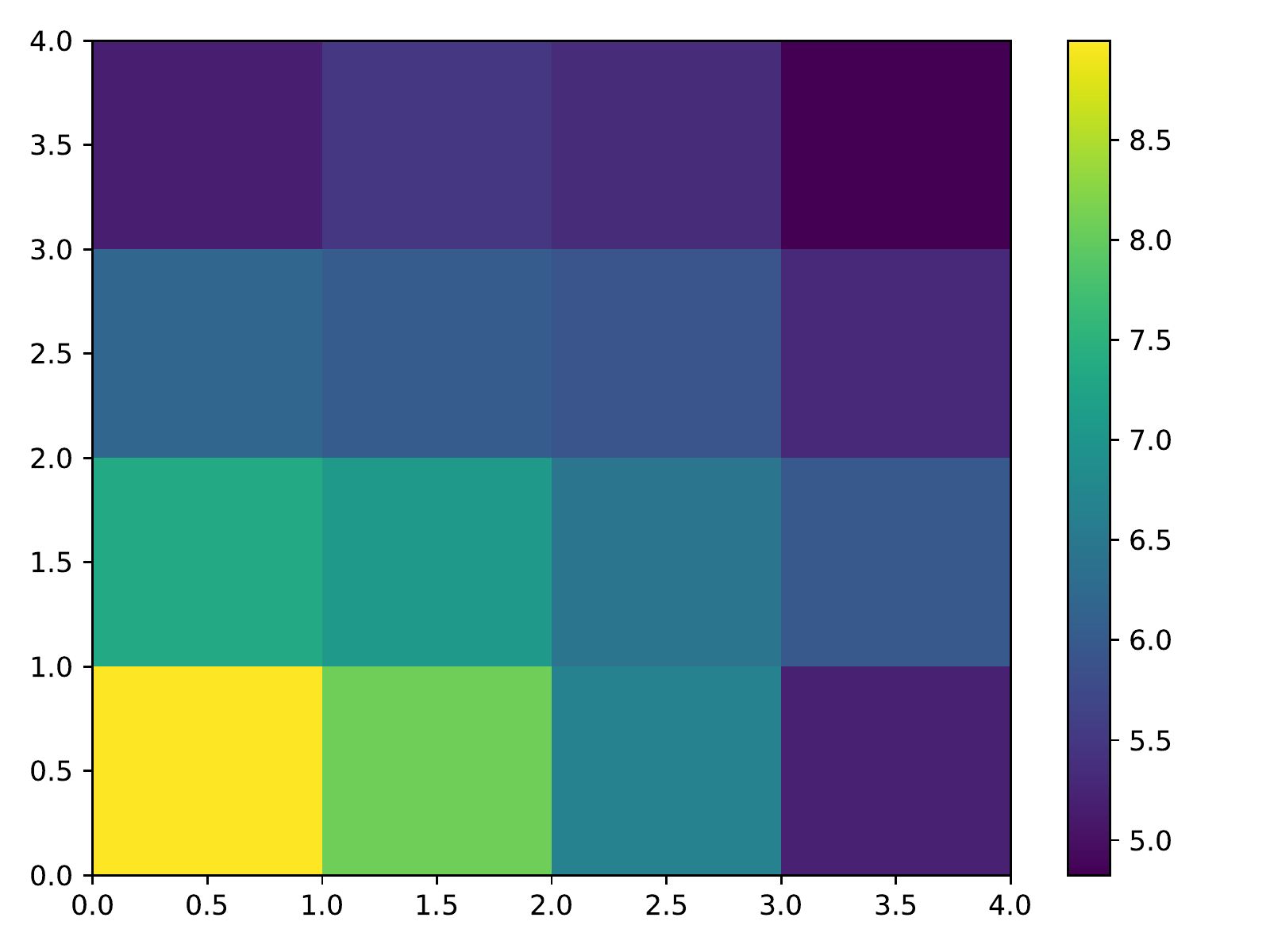}
\hspace{-0.15in}
\includegraphics[width=4.8cm]{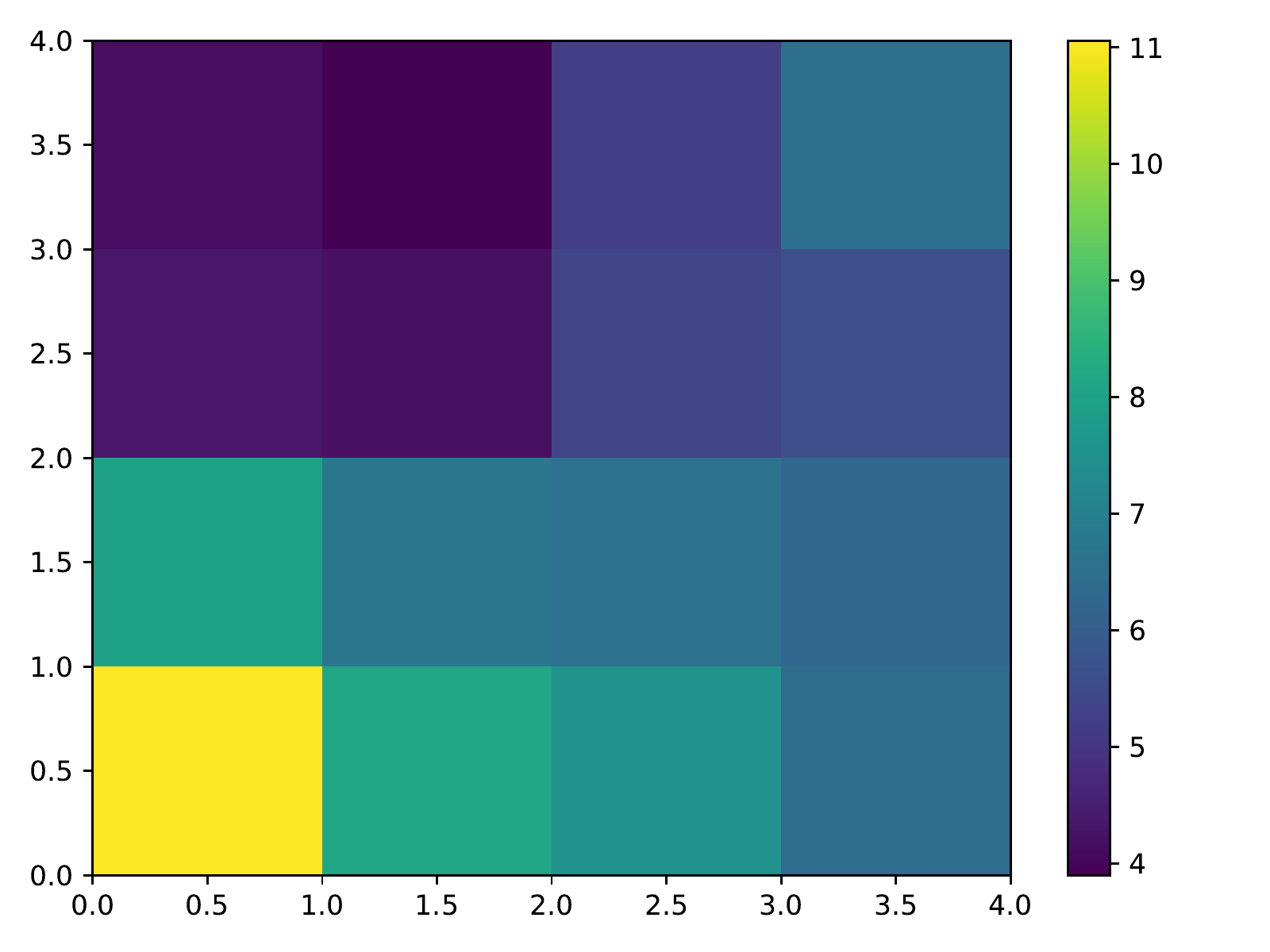}
\hspace{-0.15in}
\includegraphics[width=4.8cm]{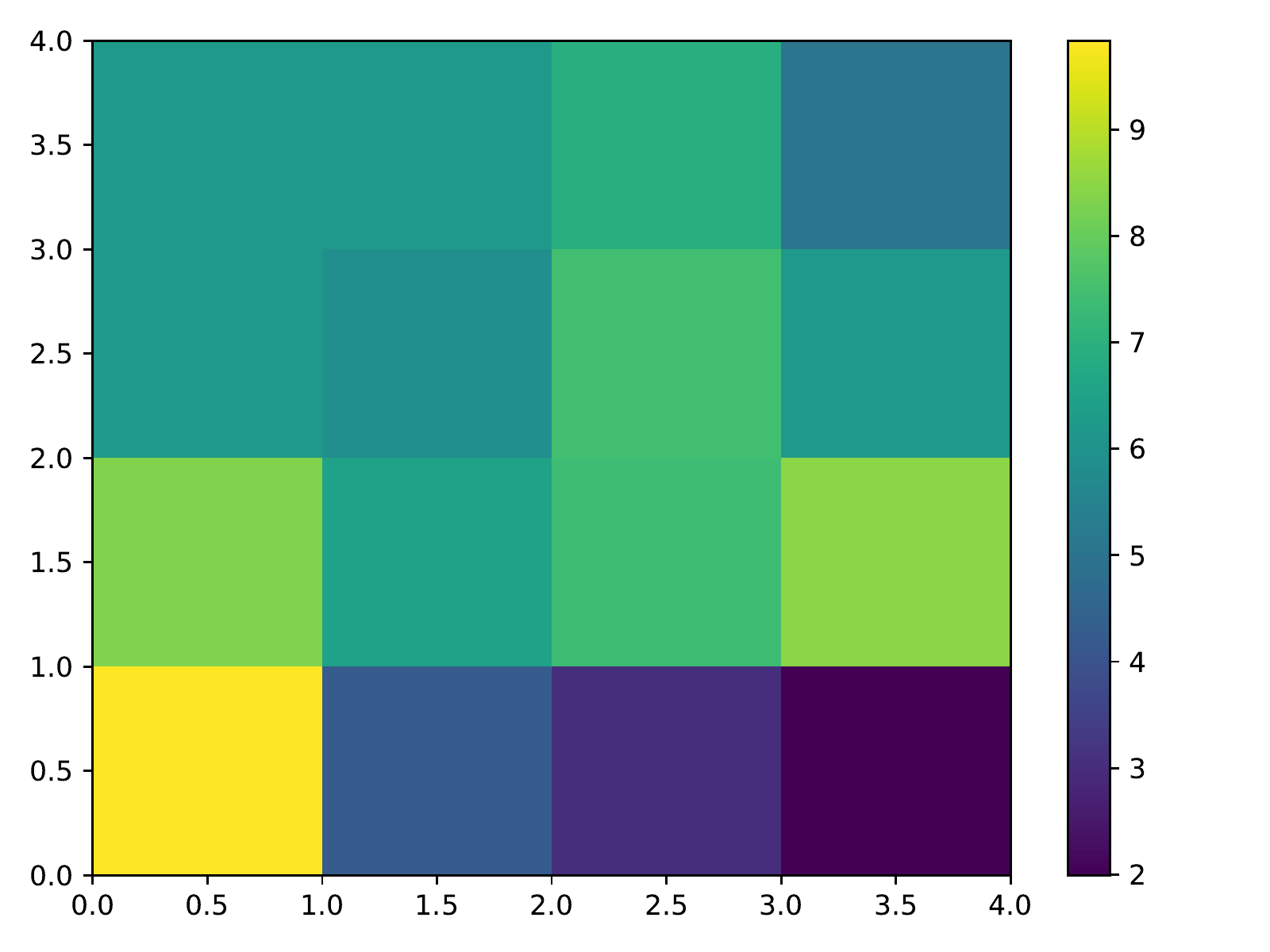}\\
\vspace{-2mm}
\hspace{-0.5cm} (a)   \hspace{3.5cm} (b) \hspace{4cm}(c) \hspace{3.5cm} (d)
\end{tabular}
\caption{Heatmap showing the percentage of state visits by the agents (aHalf=tHalf=100) for (a)~1 agent; (b)~5 agents; Heatmap showing the percentage of state visits by one agent (tHalf=100) for (c)~aHalf=50; (d)~aHalf=300.  
Compare these heatmaps with the data variance heatmap in Fig.~\ref{fig:marlcarnegie}(a).}\label{fig:heatmap_agents}
\vspace{-3mm}
\end{figure*}

\textit{Percentage of visits for each state:}
While running the MARL algorithm, we have kept a record of the number of visits for each state (by all agents) to understand where the agents spend most of the time. It is expected that the agents generally spend more time in states with high variance and vice-versa. Hence, a heatmap of these values should ideally be more close to the heatmap in Fig.~\ref{fig:marlcarnegie}(a). We have plotted these heatmap values with $aHalf=tHalf=100$ for one agent and five agents in Fig.~\ref{fig:heatmap_agents}(a) - (b) for comparison purpose. These two cases indicate a small and medium number of agents respectively. We can notice that Fig.~\ref{fig:heatmap_agents}(a) closely resembles the reference heatmap, which indicates that a less number of agents is preferred. We have plotted similar heatmap (percentage of state visits), with one agent, $tHalf=100$ for variation in aHalf (low, medium) in Fig.~\ref{fig:heatmap_agents}(c) - (d).
From these results, we can observe that Fig.~\ref{fig:heatmap_agents}(d) closely resembles the reference heatmap and hence a medium value for $aHalf$ is preferred. Similar result for variation in tHalf (one agent, $aHalf=100$) 
tells that a medium value for $tHalf$ (=$300$) is preferred. In particular, we observed a low $tHalf=50$ fares badly in that the most visited state is not the state with highest variance (lower left). Analyzing all these results (number of steps needed for policy convergence and percentage of state visits), we can conclude that the following parameters are optimum for our setting: number of agents less than 10, aHalf and tHalf around 200. Although the algorithm is scalable, all agents must be able to communicate with the server and thus must be in range of the control station; this issue imposes limitation on the optimal number of agents.

\section{Conclusion and Future Work}\label{sec:conc}
We proposed a Multi-Agent Reinforcement Learning~(MARL) framework to make efficient sequence of decisions for underwater adaptive sampling using autonomous vehicles. 
The solution was evaluated via computer simulations (to find optimal values for different design parameters) as well as field experiments, and was shown to achieve the desired performance. 
%
As future work, we plan to increase the precision in terms of accounting for the currents and drifts. 
We will also consider dynamic aspects of the environment, in which the measurements from the sensors may be noisy. 
\textbf{Acknowledgment:}
This work was supported by the NSF CPS Award No.~1739315. We thank Agam Modasiya and Karun Kanda (Rutgers MAE and CS students) for their help with the experiments.

\bibliographystyle{ACM-Reference-Format}
\bibliography{ref,refs-vidya}


\begin{thebibliography}{18}


\ifx \showCODEN    \undefined \def \showCODEN     #1{\unskip}     \fi
\ifx \showDOI      \undefined \def \showDOI       #1{#1}\fi
\ifx \showISBNx    \undefined \def \showISBNx     #1{\unskip}     \fi
\ifx \showISBNxiii \undefined \def \showISBNxiii  #1{\unskip}     \fi
\ifx \showISSN     \undefined \def \showISSN      #1{\unskip}     \fi
\ifx \showLCCN     \undefined \def \showLCCN      #1{\unskip}     \fi
\ifx \shownote     \undefined \def \shownote      #1{#1}          \fi
\ifx \showarticletitle \undefined \def \showarticletitle #1{#1}   \fi
\ifx \showURL      \undefined \def \showURL       {\relax}        \fi
\providecommand\bibfield[2]{#2}
\providecommand\bibinfo[2]{#2}
\providecommand\natexlab[1]{#1}
\providecommand\showeprint[2][]{arXiv:#2}

\bibitem[\protect\citeauthoryear{Chen, Pandey, and Pompili}{Chen
  et~al\mbox{.}}{2012}]%
        {chen2012distributed}
\bibfield{author}{\bibinfo{person}{Baozhi Chen}, \bibinfo{person}{Parul
  Pandey}, {and} \bibinfo{person}{Dario Pompili}.}
  \bibinfo{year}{2012}\natexlab{}.
\newblock \showarticletitle{A distributed adaptive sampling soluting using
  autonomous underwater vehicles}. In \bibinfo{booktitle}{\emph{Proceedings of
  the ACM International Conference on Underwater Networks and Systems
  (Wuwnet)}}. ACM, \bibinfo{pages}{29--36}.
\newblock


\bibitem[\protect\citeauthoryear{Huang, Yang, and Liu}{Huang
  et~al\mbox{.}}{2005}]%
        {distqlearn3}
\bibfield{author}{\bibinfo{person}{Jing Huang}, \bibinfo{person}{Bo Yang},
  {and} \bibinfo{person}{Da-you Liu}.} \bibinfo{year}{2005}\natexlab{}.
\newblock \showarticletitle{{A Distributed Q-Learning Algorithm for Multi-Agent
  Team Coordination}}. In \bibinfo{booktitle}{\emph{Machine Learning and
  Cybernetics, 2005. Proceedings of 2005 International Conference on}},
  Vol.~\bibinfo{volume}{1}. \bibinfo{pages}{108--113}.
\newblock
\urldef\tempurl%
\url{https://doi.org/10.1109/ICMLC.2005.1526928}
\showDOI{\tempurl}


\bibitem[\protect\citeauthoryear{La, Lim, and Sheng}{La et~al\mbox{.}}{2014}]%
        {lalim2015}
\bibfield{author}{\bibinfo{person}{Hung~Manh La}, \bibinfo{person}{Ronny Lim},
  {and} \bibinfo{person}{Weihua Sheng}.} \bibinfo{year}{2014}\natexlab{}.
\newblock \showarticletitle{Multirobot Cooperative Learning for Predator
  Avoidance}.
\newblock \bibinfo{journal}{\emph{IEEE Transactions on Control Systems
  Technology}} \bibinfo{volume}{23}, \bibinfo{number}{1}
  (\bibinfo{year}{2014}), \bibinfo{pages}{52--63}.
\newblock


\bibitem[\protect\citeauthoryear{Mariano and Morales}{Mariano and
  Morales}{2000}]%
        {distqlearn2}
\bibfield{author}{\bibinfo{person}{Carlos Mariano} {and}
  \bibinfo{person}{Eduardo Morales}.} \bibinfo{year}{2000}\natexlab{}.
\newblock \showarticletitle{{A New Distributed Reinforcement Learning Algorithm
  for Multiple Objective Optimization Problems}}.
\newblock In \bibinfo{booktitle}{\emph{Advances in Artificial Intelligence}},
  \bibfield{editor}{\bibinfo{person}{MariaCarolina Monard} {and}
  \bibinfo{person}{JaimeSim{\~{A}}{\pounds}o Sichman}} (Eds.).
  \bibinfo{series}{Lecture Notes in Computer Science},
  Vol.~\bibinfo{volume}{1952}. \bibinfo{publisher}{Springer Berlin Heidelberg},
  \bibinfo{pages}{290--299}.
\newblock
\showISBNx{978-3-540-41276-2}
\urldef\tempurl%
\url{https://doi.org/10.1007/3-540-44399-1_30}
\showDOI{\tempurl}


\bibitem[\protect\citeauthoryear{Megherbi and Malayia}{Megherbi and
  Malayia}{2012}]%
        {megherbi2012}
\bibfield{author}{\bibinfo{person}{Dalila~B. Megherbi} {and}
  \bibinfo{person}{Vikram Malayia}.} \bibinfo{year}{2012}\natexlab{}.
\newblock \showarticletitle{Cooperation in a distributed hybrid
  potential-field/reinforcement learning multi-agents-based autonomous path
  planning in a dynamic time-varying unstructured environment}. In
  \bibinfo{booktitle}{\emph{2012 IEEE International Multi-Disciplinary
  Conference on Cognitive Methods in Situation Awareness and Decision
  Support}}. \bibinfo{pages}{80--87}.
\newblock
\showISSN{2379-1667}
\urldef\tempurl%
\url{https://doi.org/10.1109/CogSIMA.2012.6188413}
\showDOI{\tempurl}


\bibitem[\protect\citeauthoryear{Rahmati, Karten, and Pompili}{Rahmati
  et~al\mbox{.}}{2018}]%
        {rahmatislam2018}
\bibfield{author}{\bibinfo{person}{Mehdi Rahmati}, \bibinfo{person}{Seth
  Karten}, {and} \bibinfo{person}{Dario Pompili}.}
  \bibinfo{year}{2018}\natexlab{}.
\newblock \showarticletitle{SLAM-based Underwater Adaptive Sampling Using
  Autonomous Vehicles}. In \bibinfo{booktitle}{\emph{Proceedings of MTS/IEEE
  OCEANS}}. IEEE, \bibinfo{pages}{1--7}.
\newblock


\bibitem[\protect\citeauthoryear{Rahmati and Pompili}{Rahmati and
  Pompili}{2017a}]%
        {rahmati2017ssfb}
\bibfield{author}{\bibinfo{person}{Mehdi Rahmati} {and} \bibinfo{person}{Dario
  Pompili}.} \bibinfo{year}{2017}\natexlab{a}.
\newblock \showarticletitle{SSFB: Signal-Space-Frequency Beamforming for
  Underwater Acoustic Video Transmission}. In
  \bibinfo{booktitle}{\emph{Proceedings of the International Conference on
  Mobile Ad Hoc and Sensor Systems (MASS)}}. IEEE, \bibinfo{pages}{180--188}.
\newblock


\bibitem[\protect\citeauthoryear{Rahmati and Pompili}{Rahmati and
  Pompili}{2017b}]%
        {rahmati2017unisec}
\bibfield{author}{\bibinfo{person}{Mehdi Rahmati} {and} \bibinfo{person}{Dario
  Pompili}.} \bibinfo{year}{2017}\natexlab{b}.
\newblock \showarticletitle{UNISeC: Inspection, separation, and classification
  of underwater acoustic noise point sources}.
\newblock \bibinfo{journal}{\emph{IEEE Journal of Oceanic Engineering}}
  \bibinfo{volume}{43}, \bibinfo{number}{3} (\bibinfo{year}{2017}),
  \bibinfo{pages}{777--791}.
\newblock


\bibitem[\protect\citeauthoryear{Rahmati and Pompili}{Rahmati and
  Pompili}{2018}]%
        {rahmati2018probabilistic}
\bibfield{author}{\bibinfo{person}{Mehdi Rahmati} {and} \bibinfo{person}{Dario
  Pompili}.} \bibinfo{year}{2018}\natexlab{}.
\newblock \showarticletitle{Probabilistic spatially-divided multiple access in
  underwater acoustic sparse networks}.
\newblock \bibinfo{journal}{\emph{IEEE Transactions on Mobile Computing}}
  (\bibinfo{year}{2018}).
\newblock
\urldef\tempurl%
\url{https://doi.org/10.1109/TMC.2018.2877683}
\showDOI{\tempurl}


\bibitem[\protect\citeauthoryear{Sadhu, Devaraj, and Pompili}{Sadhu
  et~al\mbox{.}}{2018}]%
        {Sadhu2018ucomms}
\bibfield{author}{\bibinfo{person}{Vidyasagar Sadhu}, \bibinfo{person}{Sanjana
  Devaraj}, {and} \bibinfo{person}{Dario Pompili}.}
  \bibinfo{year}{2018}\natexlab{}.
\newblock \showarticletitle{{Energy-efficient Wireless Analog Sensing for
  Persistent Underwater Environmental Monitoring}}. In
  \bibinfo{booktitle}{\emph{2018 IEEE Third Underwater Communications and
  Networking Conference (UComms)}}. \bibinfo{pages}{1--4}.
\newblock


\bibitem[\protect\citeauthoryear{Sadhu, Pompili, Zonouz, and Sritapan}{Sadhu
  et~al\mbox{.}}{2017}]%
        {Sadhu2017icccn}
\bibfield{author}{\bibinfo{person}{Vidyasagar Sadhu}, \bibinfo{person}{Dario
  Pompili}, \bibinfo{person}{Saman Zonouz}, {and} \bibinfo{person}{Vincent
  Sritapan}.} \bibinfo{year}{2017}\natexlab{}.
\newblock \showarticletitle{{CollabLoc: Privacy-preserving multi-modal
  localization via collaborative information fusion}}. In
  \bibinfo{booktitle}{\emph{Proc. of the Intl. Conference on Computer
  Communications and Networks (ICCCN)}}. \bibinfo{publisher}{IEEE},
  \bibinfo{address}{Vancouver, BC}.
\newblock


\bibitem[\protect\citeauthoryear{Sadhu, Salles-Loustau, Pompili, Zonouz, and
  Sritapan}{Sadhu et~al\mbox{.}}{2016}]%
        {Sadhu2016icac}
\bibfield{author}{\bibinfo{person}{Vidyasagar Sadhu}, \bibinfo{person}{Gabriel
  Salles-Loustau}, \bibinfo{person}{Dario Pompili}, \bibinfo{person}{Saman
  Zonouz}, {and} \bibinfo{person}{Vincent Sritapan}.}
  \bibinfo{year}{2016}\natexlab{}.
\newblock \showarticletitle{{Argus: Smartphone-enabled human cooperation via
  multi-agent reinforcement learning for disaster situational awareness}}. In
  \bibinfo{booktitle}{\emph{Proc. of IEEE Intl. Conference on Autonomic
  Computing, ICAC}}.
\newblock


\bibitem[\protect\citeauthoryear{Sartoretti, Shi, Paivine, Travers, and
  Choset}{Sartoretti et~al\mbox{.}}{2018}]%
        {Sartoretti2018}
\bibfield{author}{\bibinfo{person}{Guillaume Sartoretti},
  \bibinfo{person}{Yunfei Shi}, \bibinfo{person}{William Paivine},
  \bibinfo{person}{Matthew Travers}, {and} \bibinfo{person}{Howie Choset}.}
  \bibinfo{year}{2018}\natexlab{}.
\newblock \showarticletitle{Distributed Learning for the Decentralized Control
  of Articulated Mobile Robots}. In \bibinfo{booktitle}{\emph{Procedings of
  IEEE International Conference on Robotics and Automation (ICRA)}}.
  \bibinfo{pages}{1--6}.
\newblock
\showISSN{2577-087X}
\urldef\tempurl%
\url{https://doi.org/10.1109/ICRA.2018.8460802}
\showDOI{\tempurl}


\bibitem[\protect\citeauthoryear{Tan}{Tan}{1993}]%
        {tan1993multi}
\bibfield{author}{\bibinfo{person}{Ming Tan}.} \bibinfo{year}{1993}\natexlab{}.
\newblock \showarticletitle{Multi-agent reinforcement learning: Independent vs.
  cooperative agents}. In \bibinfo{booktitle}{\emph{Proceedings of the
  International Conference on Machine Learning}}. \bibinfo{pages}{330--337}.
\newblock


\bibitem[\protect\citeauthoryear{Wang, Wei, Wang, Song, and Mahmoudian}{Wang
  et~al\mbox{.}}{2018}]%
        {wang2018}
\bibfield{author}{\bibinfo{person}{Chaofeng Wang}, \bibinfo{person}{Li Wei},
  \bibinfo{person}{Zhaohui Wang}, \bibinfo{person}{Min Song}, {and}
  \bibinfo{person}{Nina Mahmoudian}.} \bibinfo{year}{2018}\natexlab{}.
\newblock \showarticletitle{Reinforcement Learning-Based Multi-AUV Adaptive
  Trajectory Planning for Under-Ice Field Estimation}.
\newblock \bibinfo{journal}{\emph{Sensors}}  \bibinfo{volume}{18}
  (\bibinfo{date}{11} \bibinfo{year}{2018}), \bibinfo{pages}{3859}.
\newblock
\urldef\tempurl%
\url{https://doi.org/10.3390/s18113859}
\showDOI{\tempurl}


\bibitem[\protect\citeauthoryear{Yu~Fan~Chen}{Yu~Fan~Chen}{2017}]%
        {chenliu2017icra}
\bibfield{author}{\bibinfo{person}{Michael Everett Jonathan P.~How Yu~Fan~Chen,
  Miao~Liu}.} \bibinfo{year}{2017}\natexlab{}.
\newblock \showarticletitle{Decentralized non-communicating multiagent
  collision avoidance with deep reinforcement learning}. In
  \bibinfo{booktitle}{\emph{Proceedings of IEEE International Conference on
  Robotics and Automation (ICRA)}}. \bibinfo{pages}{285--292}.
\newblock


\bibitem[\protect\citeauthoryear{Zhao, Sadhu, Yang, and Pompili}{Zhao
  et~al\mbox{.}}{2018}]%
        {Zhao2018a}
\bibfield{author}{\bibinfo{person}{Xueyuan Zhao}, \bibinfo{person}{Vidyasagar
  Sadhu}, \bibinfo{person}{Anthony Yang}, {and} \bibinfo{person}{Dario
  Pompili}.} \bibinfo{year}{2018}\natexlab{}.
\newblock \showarticletitle{{Improved Circuit Design of Analog Joint Source
  Channel Coding for Low-Power and Low-Complexity Wireless Sensors}}.
\newblock \bibinfo{journal}{\emph{IEEE Sensors Journal}} \bibinfo{volume}{18},
  \bibinfo{number}{1} (\bibinfo{date}{jan} \bibinfo{year}{2018}),
  \bibinfo{pages}{281--289}.
\newblock


\bibitem[\protect\citeauthoryear{Zhou and Shen}{Zhou and Shen}{2011}]%
        {zhoushen2011}
\bibfield{author}{\bibinfo{person}{Pucheng Zhou} {and} \bibinfo{person}{Huiyan
  Shen}.} \bibinfo{year}{2011}\natexlab{}.
\newblock \showarticletitle{Multi-agent cooperation by reinforcement learning
  with teammate modeling and reward allotment}. In
  \bibinfo{booktitle}{\emph{Proceedings of International Conference on Fuzzy
  Systems and Knowledge Discovery (FSKD)}}, Vol.~\bibinfo{volume}{2}. IEEE,
  \bibinfo{pages}{1316--1319}.
\newblock


\end{thebibliography}
\end{document}